\documentclass[aps,prl,twocolumn,floatfix,superscriptaddress,amsmath]{revtex4-2}

\usepackage[unicode=true,colorlinks=true]{hyperref}
\usepackage{graphicx}
\usepackage{bm}
\usepackage{bbold}
\usepackage{mathrsfs}

\newcommand{\KP}{{\bf k}\ensuremath{\cdot}{\bf p}}

\newcommand{\ket}[1]{\left\vert #1 \right\rangle}

\newcommand{\rmi}{{\rm i}}

\begin{document}

\title{Effective Land{\'e} factors of electrons and holes in lead chalcogenide nanocrystals}
\author{I.D.~Avdeev} 
\affiliation{Ioffe Institute, 194021 St.\,Petersburg, Russia}
\author{S.V.~Goupalov}
\email{serguei.goupalov@jsums.edu}
\affiliation{Ioffe Institute, 194021 St.\,Petersburg, Russia}
\affiliation{Department of Physics, Jackson State University, Jackson MS 39217, USA}
\author{M.O.~Nestoklon}
\affiliation{Ioffe Institute, 194021 St.\,Petersburg, Russia}




\begin{abstract}
  The Land{\'e} or $g$-factors of charge carriers in solid state systems provide invaluable information about response of quantum states to external magnetic fields and are key ingredients in description of spin-dependent phenomena in nanostructures.
We report on the comprehensive theoretical analysis of electron and hole $g$-factors in lead chalcogenide nanocrystals.
By combining symmetry analysis, atomistic calculations, and extended \KP\ theory, we relate calculated linear-in-magnetic field energy splittings of confined electron states in nanocrystals to the intravalley $g$-factors of the multi-valley bulk materials, renormalized due to the quantum confinement. We demonstrate that this renormalization is correctly reproduced by analytical expressions derived in the framework of the extended \KP\ model. 
\end{abstract}

\maketitle

{\it Introduction.}
Lead salts nanocrystals (NCs) are enjoying many practical applications in optoelectronics and photovoltaics~\cite{Sun2012,Sukhovatkin09,Tisdale2010,Gao2020,Lu2020}. New devices built on NCs are predicted to enter the market in the nearest future~\cite{konstantatos,kagan,kirmani,Lu2020}. All these devices are based on the emission or absorption of light by spatially confined electron-hole pairs.

Applications of NCs in rapidly developing fields of spintronics
and quantum computing \cite{Loss98,Imamoglu99,Warburton13,Cao16} would be impossible without control over the spin state of localized carriers. 
Therefore,  knowledge about carrier spin relaxation and dynamics as well as their
Land\'e $g$ factors becomes critically important. These properties have been widely studied for CdSe NCs. The exciton fine structure relaxation dynamics was investigated in Refs.~\citenum{scholes1,scholes2,scholes3}, electron and exciton $g$-factors were measured, respectively, by the time-resolved Faraday rotation~\cite{Zhang2014,Hu2019,Wu2021} and 
single-dot magneto-photoluminescence spectroscopy~\cite{fernee,Sinito14}, and
carrier $g$-factors were calculated within tight-binding~\cite{schrier03,Chen04,Tadjine17} and effective mass~\cite{Semina21} methods.

In the mean time, analogous studies for lead salts NCs remain very scarce. Ultrafast exciton fine structure relaxation dynamics was studied by Johnson et al.~\cite{Johnson08} Schaller et al. measured averaged exciton $g$-factor in an ensemble of PbSe NCs in magnetic-circular dichroism experiments~\cite{Schaller10}. Turyanska et al. deduced exciton $g$-factors of PbS NCs from magnetic field dependences of photoluminescence circular polarization degree~\cite{Turyanska10}. Single-NC spectroscopy in external magnetic fields was performed by Kim et al~\cite{Kim21}.
\begin{figure*}[tb]
\includegraphics[width=0.9\textwidth]{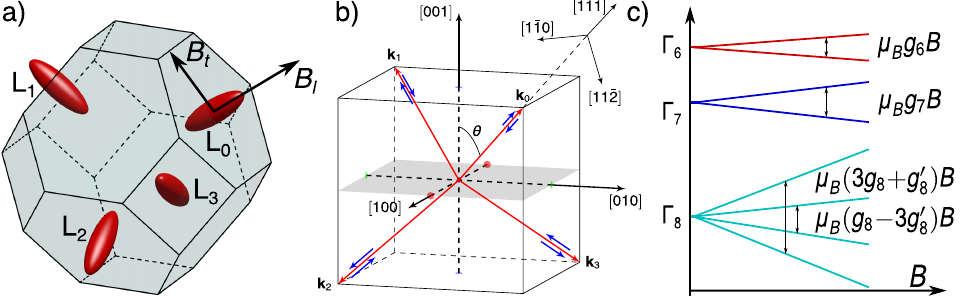}
\caption{
(a) Conduction and valence band extrema in bulk lead chalcogenide compounds occur at four ineqivalent $L$-points of the Brillouin zone. Energy isosurfaces near these points form anisotropic valleys (red ``cigars''). For valley states of charge carriers, Zeeman splittings depend on the orientation of the magnetic field with respect to the valley main axes. One can distinguish longitudinal ($B_l$) and transverse ($B_t$) components
of the magnetic field in a given valley which define the longitudinal ($g_l$) and transverse ($g_t$) $g$-factors. 
(b) Within the \KP\ theory, the ``spin'' (total angular momentum) projections $F_z = \pm 1/2$ (blue arrows) are defined in the valley coordinate frames adjacent to the valley wave vectors ${\bf k}_{\nu}$ (red arrows). 
These vectors form an irreducible star $\{ {\bf k}_0 \}$ of some representation of the crystal space group~\cite{BirPikus}. One can analyze their transformation properties \cite{Note1} 
and form combinations of valley states transforming under irreducible representations of the NC point group.
This allows one to establish a relationship between the single-valley states described by the \KP\ theory and multivalley combinations of states, enforced by the symmetry and following from atomistic calculations of the NC band structure.
(c) Zero-field states of charge carriers in NCs represent combinations of valley states transforming with respect to certain irreducible representations of the symmetry group. The ground electron (or hole) level in a NC with no inversion center splits into two doublets and a quadruplet transforming with respect to the $\Gamma_6$, $\Gamma_7$, and $\Gamma_8$ representations of the group $T_d$. At low magnetic fields, the Zeeman splittings are isotropic and determined by the effective $g$-factors $g_6$, $g_7$, $g_8$, and $g'_8$. 
}
\label{fig:illus}
\end{figure*}

Yet, interpretation of these results is complicated by the multi-valley band structure of lead salts compounds.
Bulk lead salts have extrema of the conduction and valence bands at the four inequivalent $L$-points of the Brillouin zone. The widely used \KP\ 
theory \cite{Kang1997} treats these $L$-valleys independently. An external magnetic field leads to the Zeeman splittings of the electron and hole 
states characterized by certain magnetic quantum numbers intimately related to the spin degrees of freedom. Then the main effects of the quantum confinement
are renormalization of the Zeeman splittings and their sensitivity to orientation of the magnetic field, which result in the renormalization and anisotropy of the carriers' $g$-factors~\cite{Kiselev98}. This kind of narrative is typical for nanostructures of II~-~VI and III~-~V compound semiconductors with band extrema at 
the $\Gamma$ point of the Brillouin zone, and is adopted by the conventional, or independent-valley,  \KP\ theory developed for lead salt nanostructures~\cite{Kang1997}. This theory is formulated in terms of the longitudinal and transverse single-valley $g$-factors (Figure~\ref{fig:illus},~(a)). 

However, in lead salts nanostructures, due to the inter-valley scattering on the surface, the zero-field electron or hole states represent combinations of the states originating from different $L$-valleys. Thus, all atomistic band structure calculations, basing on the symmetries of the underlying crystal lattice
and overall structure, 
automatically take into account this alignment of the valley degrees of freedom~\cite{Delerue2004b,Poddubny12}. The resulting zero-field states are classified with respect to irreducible representations of the symmetry group. Application of the external magnetic field further affects the spin degrees of freedom, but this narrative implies completely different meaning and definition of the magnetic quantum numbers as compared to the single-valley case. Since, at weak magnetic fields, states characterized by different irreducible representations do not mix, the atomistic theories operate with the $g$-factors associated with the corresponding irreducible representations (Figure~\ref{fig:illus},~(c)).

In this work we show that a solution to this ambiguity comes from a symmetry-based construction of a transformation relating the basis of 
independent valley states and the basis of valley combinations associated with certain irreducible representations of the point group, as illustrated in Figure~\ref{fig:illus}.
This allows one to relate both kinds of the $g$-factors and use a fusion of the two approaches to get insight about confinement effect
on carriers' $g$-factors in lead salts NCs.


{\it Results and discussion.}
In PbX (X=S, Se) NCs with cubic symmetry (point group $T_d$ or $O_h$) the ground state of confined electron or hole splits into two doublets, 
transforming under irreducible representations $\Gamma_6,\Gamma_7$ ($\Gamma_{6,7}^{\pm}$) of group $T_d$ ($O_h$), and a quadruplet $\Gamma_8$ ($\Gamma_8^{\pm}$) separated by several meV as a result of valley mixing~\cite{Delerue2004b,Poddubny12,Avdeev2020}.
In the subspace of these states, interaction with a weak magnetic field ${\bf B}$ is described by the following effective Hamiltonian,
written as a block-diagonal matrix:
\begin{equation}\label{eq:Hinv1}
{H}_1^{\eta}({\bf B}) = 
  \mu_B{\bf B}
  \begin{pmatrix}
    \frac12 g_6^{\eta} \bm \sigma & 0 & 0 \\
    0 & \frac12 g_7^{\eta} \bm \sigma & 0 \\
	  0 & 0 & g_8^{\eta} {\bf J} + {g'}_8^{\eta} {\bf J}' \\
  \end{pmatrix}\,,
\end{equation}
where $\eta=c(v)$ for the conduction (valence) band states; $g_6^{\eta}$ and $g_7^{\eta}$ are, respectively, the effective $g$-factors of the $\Gamma_6$ and $\Gamma_7$ doublets; $g_8^{\eta}$ and ${g'}_8^{\eta}$ are the two constants describing Zeeman splitting of the quadruplet $\Gamma_8$;
$\bm \sigma = (\sigma_x, \sigma_y, \sigma_z)$ are the Pauli matrices; ${\bf J} = (J_x, J_y, J_z)$ are the matrices of the angular momentum $j=3/2$,\footnote{See Supplemental Material at .... for details of the details of calculations, results of atomistic calculations ofr different QD shapes, details of calculations in isotropic and anisotropic \KP\ model and explicit form of the matrices transforming the basis functions of independent valleys to the basis of irreducible representations of the group $T_d$.} $\mu_B$ is the Bohr magneton, and the matrices ${\bf J}'$ are defined as\footnote{Note that in Ref.~\citenum{bookIvchenko95} the effective Hamiltonian is written in terms of the matrices $J_{\gamma}^3$.}
\begin{equation}\label{eq:Jp_def}
  J_{\gamma}' = \frac53 J_{\gamma}^3 - \frac{41}{12} J_{\gamma}\,
\end{equation}
$(\gamma=x,y,z)$.

In a strong magnetic field, when $\mu_BB$ is compatible with the valley splittings $|E_{\Gamma_7}-E_{\Gamma_8}|$ and $|E_{\Gamma_8}-E_{\Gamma_6} |$, two additional non-diagonal linear-in-$B$ terms should be taken into account.
They describe interaction of the quadruplet $\Gamma_8$ with the doublets $\Gamma_6$ and $\Gamma_7$ and are discussed in Supplemental Material~\cite{Note1}. 


{\it Tight-binding calculations.}
The effective $g$-factors entering equation~(\ref{eq:Hinv1}) can be extracted from the tight-binding calculations~\cite{Note1}.
\begin{figure}[tb]
\includegraphics[width=0.9\linewidth]{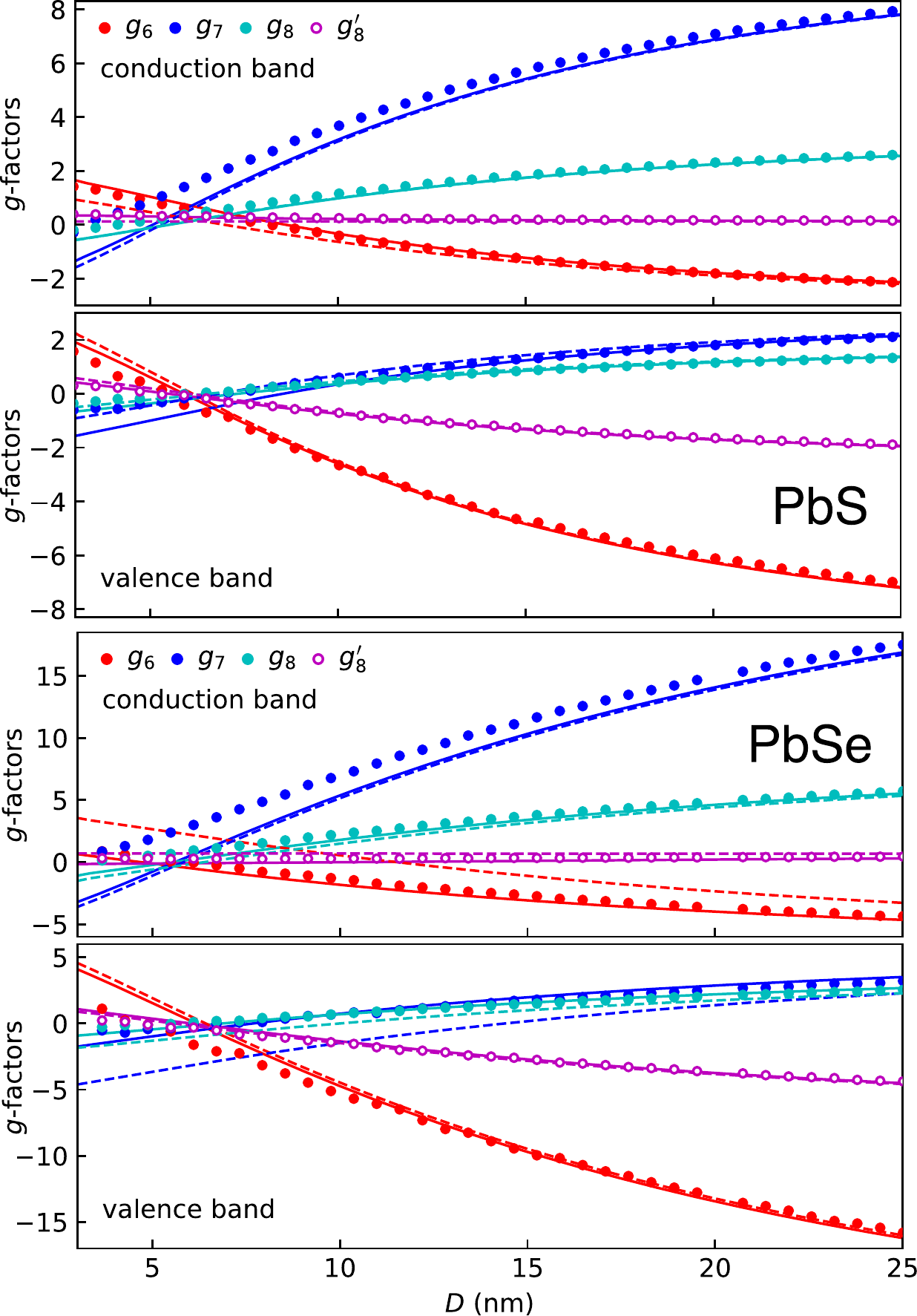}
\caption{
Calculated $g$-factors $g_6,g_7,g_8$ and $g_8'$ \eqref{eq:Hinv1} in conduction (top panel) and valence (second panel) bands of quasi-spherical PbS NCs 
(filled red, blue, cyan, and empty purple circles, respectively). The results are stable with respect to NC shape variation, see Supplemental Material~\cite{Note1}.
 Solid (dashed) lines show outcomes of the anisotropic (isotropic) \KP\ model.
 Two lower panels show results for PbSe NCs.
}
\label{fig:g678_PbS}
\end{figure}
They are shown in Figure~\ref{fig:g678_PbS} for quasi-spherical PbS and PbSe NCs
(see Supplemental Material~\cite{Note1} for definition of quai-spherical NCs).
The actual shapes of colloidal NCs can vary from cube to truncated cube to cuboctahedron to truncated octahedron to octahedron, depending on the synthesis conditions~\cite{Bealing12,Choi2013}.
Tight-binding calculations performed for NCs of cubic, cuboctahedral, and octahedral shapes show that the $g$-factors are almost shape-independent, in contrast to the zero-field splittings of electron and hole levels exhibiting strong dependencies on the NC shape~\cite{Avdeev2020}
(see Supplemetal Material~\cite{Note1} for energies of the $\Gamma_6,\Gamma_7$, and $\Gamma_8$ levels in PbSe NCs. These for PbS NCs are given in Ref. \citenum{Avdeev2020}.)
By dashed (solid) lines in Figure~\ref{fig:g678_PbS} we show the results of the isotropic (anisotropic) \KP\ model to be discussed later.


{\it Land{\'e} factors in valleys.}
Before proceeding to the results of the \KP\ model, we first discuss the origin of the parameters entering Eq.~\eqref{eq:Hinv1} in terms of
anisotropic $g$-factors describing linear magnetic field dependence of the phenomenological single-valley effective Hamiltonians.
We represent the Hamiltonian of the confined conduction- or valence-band ground state as a summation of the single-valley Hamiltonians over the valley index $\nu$:
\begin{equation}\label{eq:Hglt}
  \hat H^{\eta} = 
  \frac{\mu_B}2  \sum_{\nu} \left[ g_t^{\eta}
    ( \sigma_x B_{x,\nu} + \sigma_y  B_{y,\nu})
  + g_l^{\eta} \sigma_z B_{z,\nu} \right] \,,
\end{equation}
where the Pauli matrices $\sigma_{\gamma}$ ($\gamma=x,y,z$) are defined in the coordinate frames of the corresponding valleys
(with the $z$ axis aligned along the valley $C_3$ axis) and ${\bf B} = (B_{x,\nu},B_{y,\nu},B_{z,\nu})$ is the magnetic field written in the 
same ``local'' basis. In particular, for 
the $L_0$ valley, we choose the local basis as follows:
\begin{equation}
	{\bf n}_{x,0} \parallel [1\bar10]\,,\quad
	{\bf n}_{y,0} \parallel [11\bar2]\,,\quad\
	{\bf n}_{z,0} \parallel [111]\,.
\end{equation}
The bases of the other valleys ($\nu=1,2,3$) are related via $C_2$ rotations around the crystallographic axes of the ``laboratory'' frame ($x \parallel [100]$, $y \parallel [010]$, $z \parallel [001]$): $0\to 1$ via $C_{2z}$, $0\to 2$ via $C_{2x}$, and $0\to 3$ via $C_{2y}$ (cf. Figure~\ref{fig:illus},~(b)).

The Hamiltonian \eqref{eq:Hglt} can be transformed into the basis of irreducible representations using an appropriate transformation matrix~\cite{Note1}, 
From a comparison of the transformed Hamiltonian with Eq.~\eqref{eq:Hinv1} we obtain the following set of the $g$-factors for the confined conduction- and valence-band states:
\begin{subequations} \label{eq:g678_glt}
\begin{align}
g_6^c & = \frac{g_l^c-2g_t^c}3\,, & g_7^c & = \frac{g_l^c+2g_t^c}3 \,,\label{eq:g67c}\\
g_8^c & = \frac{g_l^c+4g_t^c}{15}\,, & g_8^{\prime c} & = 2\frac{g_l^c-g_t^c}{15}\,,\\ 
g_6^v & = \frac{g_l^v+2g_t^v}3\,,& g_7^v & = \frac{g_l^v-2g_t^v}3 \,, \label{eq:g67v}\\
g_8^v & = \frac{g_l^v-4g_t^v}{15}\,,& g_8^{\prime v} & = 2\frac{g_l^v+g_t^v}{15}\,.
\end{align}
\end{subequations}

\begin{figure}[tb]
\includegraphics[width=0.9\linewidth]{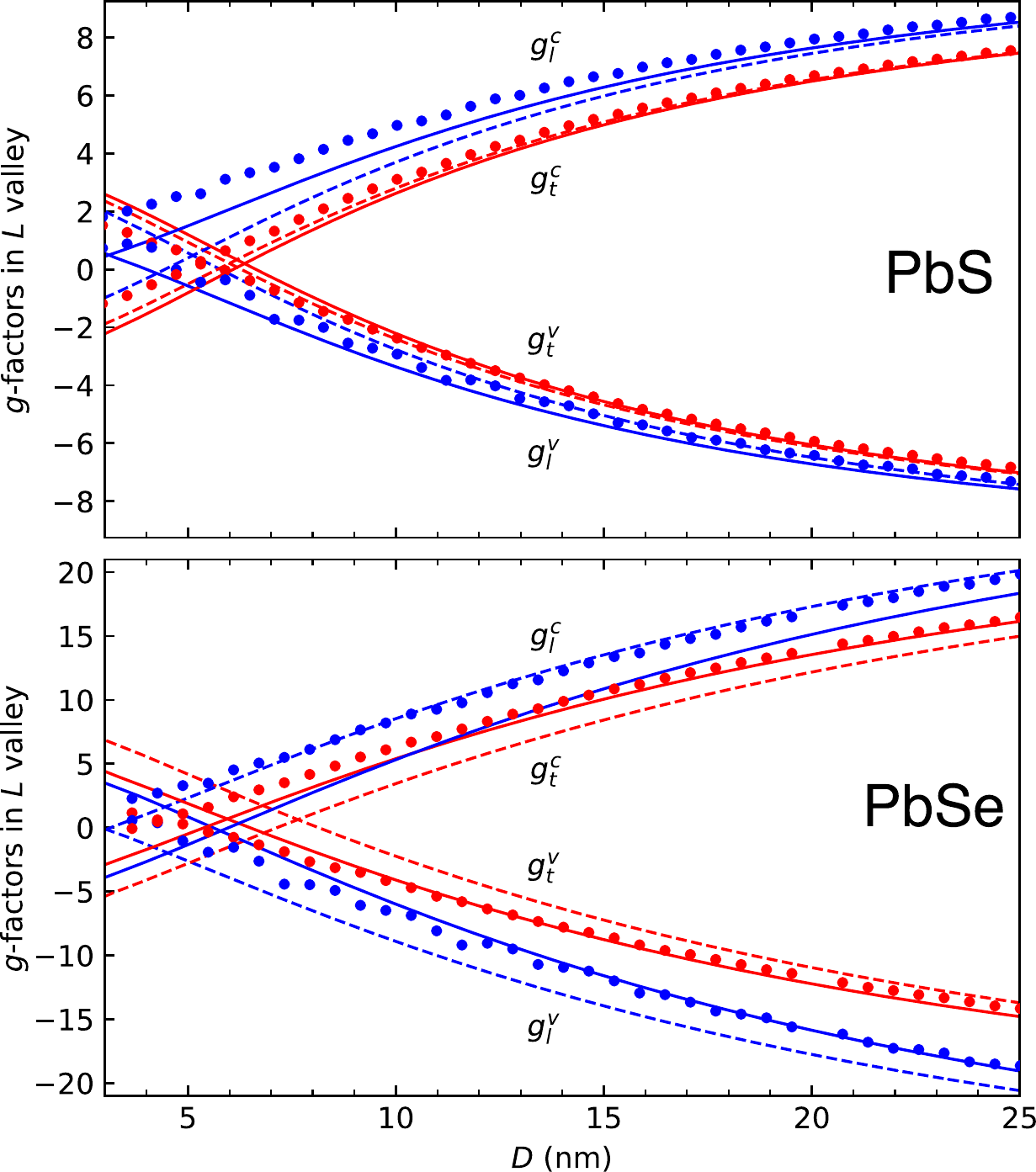}
\caption{
  The values of transverse (red circles) and longitudinal (blue circles) $L$ valley $g$-factors in conduction (positive for large $D$) and valence (negative for large $D$) bands in PbS (upper panel) and PbSe (lower panel) NCs extracted from the tight-binding calculations using Eq.~\eqref{eq:glt_tb}.  
  Red and blue lines show, respectively, longitudinal and transverse $g$-factors computed within anisotropic (solid lines) and isotropic (dashed lines) \KP\ theory with parameters extracted from the tight-binding model, Table~\ref{tb:kp_params}.
}
\label{fig:g_lt}
\end{figure}

Equations~\eqref{eq:g678_glt} can be inverted to extract the values of $g_{l(t)}^{\eta}$ from the $g$-factors of the quantum confined states in a NC.
We will use this procedure to obtain the effective $g$-factors $g_{l(t)}^{c(v)}$ from the tight-binding results presented in Figure~\ref{fig:g678_PbS}. 
One may notice that there are four independent constants for each band in the effective Hamiltonian \eqref{eq:Hinv1} but only two independent constants entering \eqref{eq:Hglt}. Therefore, we have some freedom in the choice of the extraction procedure. In the present study, we will determine the longitudinal and transverse $g$-factors as 
\begin{subequations}\label{eq:glt_tb}
\begin{align}
 g_l^{\eta} & = 3g_8^{\eta}+6{g_8^{\prime\eta}} \label{eq:gl_tb} \,,\\
 g_t^{c} & = 3g_8^{c}-\frac32{g_8^{\prime c}}\,,&
 g_t^{v} & =-3g_8^{v}+\frac32{g_8^{\prime v}}\label{eq:gt_tb} \,.
\end{align}
\end{subequations}
The longitudinal and transverse $g$-factors obtained in this manner from the tight-binding calculations are presented in Figure~\ref{fig:g_lt}. 
Also shown in Figure~\ref{fig:g_lt} are results of the single-valley \KP\ model. 

As the numbers of independent constants in equations~\eqref{eq:Hinv1} and~\eqref{eq:Hglt} are different, the above procedure introduces some error. In order to estimate it, in Supplemental Material~\cite{Note1} we compare the difference between $g_6^{\eta}$ and $g_7^{\eta}$ calculated directly in the tight-binding with the results of Eqs.~(\ref{eq:g67c},\ref{eq:g67v}). The differences are correlated with the valley splittings and do not exceed 10\% of the values of the $g$-factors in the bulk.

{\it \KP\ model.}
It is well known~\cite{Kiselev98} that quantum confinement renormalizes electron $g$-factor in semiconductor nanostructures with respect to its bulk counterpart. We will demonstrate the results of such renormalization using the single-valley \KP\ model of Ref.~\citenum{Kang1997} (see also Refs.~\citenum{Nestoklon22,Goupalov22}). 
In the isotropic approximation, the effective Hamiltonian of this model can be written as
\begin{equation}
  \label{eq:H0_iso}
  H_{\text{iso}} =
  \begin{pmatrix}
    \left(\frac{E_g}2 - \alpha_c \Delta\right) & -\frac{i\hbar P}{m_0} (\bm\sigma\bm\nabla) \\
    -\frac{i\hbar P}{m_0} (\bm\sigma\bm\nabla) & -\left(\frac{E_g}2 -\alpha_v\Delta\right)
  \end{pmatrix}
  \,,
\end{equation}
where $E_g$ is the band gap, $P$ is the inter-band momentum matrix element, $m_0$ is the free electron mass, the coefficients $\alpha_{c(v)}$
stem from the contributions of the remote bands to the conduction and valence bands energy dispersion, and $\Delta$ is the three-dimensional Laplace operator. 

Within the isotropic model, the electron states can be characterized by the value of the total angular momentum $F$ and an additional quantum number $p$ related to parity of the states. The dispersion equation for these states is given in Supplemental Material~\cite{Note1}.
The ground electron states are characterized by the angular momentum $F=1/2$ and $p=+1$ (odd parity), the ground hole states have $F=1/2$ and $p=-1$ (even parity).
All the confined states are $(2F+1)$-fold degenerate with respect to the projection $F_z$ of the total angular momentum.
Their wave functions written in the bispinor form are
\begin{equation}
  \label{eq:QD_states}
  \ket{F,p,n,F_z} = 
  \begin{pmatrix}
      f_{F-\frac{p}2,p}\left( \frac{r}{R} \right) \hat{\Omega}^{F-\frac{p}2}_{F,F_z} \\
      \cr
   \rmi \, p \, g_{F+\frac{p}2,p}\left( \frac{r}{R} \right) \hat{\Omega}^{F+\frac{p}2}_{F,F_z} 
  \end{pmatrix}\,,
\end{equation}
where $\hat{\Omega}_{F,F_z}^{\ell}$ are the spherical spinors \cite{Varshalovich} and $f_{\ell p}$, $g_{\ell p}$ are the normalized radial functions~\cite{Note1}.



To compute the $g$-factors we 
follow Ref.~\citenum{Kiselev98} and add to the Hamiltonian \eqref{eq:H0_iso} the following term
\begin{multline}\label{ham}
   \delta H=\frac{e}{2c} ({\bf Av}+{\bf vA})
\\
=\frac{e {\bf B}}{2c} 
\left(
\begin{matrix}
-\frac{2 i \alpha_c m_0}{\hbar^2} \, {\bf r} \times {\bm \nabla}
&
\frac{P}{\hbar} \, ({\bf r} \times {\bm \sigma})
\cr
\cr
\frac{P}{\hbar} \, ({\bf r} \times {\bm \sigma})
&
\frac{2 i \alpha_v m_0}{\hbar^2} \, {\bf r} \times {\bm \nabla}
\cr
\end{matrix}
\right) 
\,,
\end{multline}
where ${\bf v}$ is the velocity operator~\cite{Goupalov22}. We used the symmetric gauge ${\bf A} = \frac{{\bf  B} \times {\bf r}}2$
for the vector potential ${\bf A}$
along with the additional electron and hole effective $g$-factor tensors with the non-zero components $g_{0 \, xx}^{\eta}=g_{0 \, yy}^{\eta}= g_{0t}^{\eta}$, $g_{0 \, zz}^{\eta}= g_{0l}^{\eta}$, responsible for the contributions of remote bands, to yield
\begin{multline}
  \label{eq:HB_iso}
  H_{\text{iso}}({\bf B})\\ =
  \mu_B {\bf B}
  \begin{pmatrix}
   \frac{1}{2} \hat g_{0}^c {\bm \sigma} + \frac{2m_0  }{\hbar^2} \alpha_c{\bf L} & 
    \frac{P}{\hbar}({\bf r} \times \bm \sigma) \\
    \cr
    \frac{P}{\hbar}({\bf r} \times \bm \sigma) & 
\frac{1}{2}   \hat g_{0}^v {\bm \sigma} - \frac{2m_0 }{\hbar^2} \alpha_v{\bf L}
  \end{pmatrix}
  \,,
\end{multline}
where ${\bf L} = -\rmi {\bf r} \times \bm \nabla$ and ${\bf B} \hat g_{0}^{\eta} {\bm \sigma} =\sum\limits_{\alpha, \beta} B_{\alpha} g_{0 \, \alpha \beta}^{\eta} \sigma_{\beta}=\sum\limits_{\alpha} B_{\alpha} g_{0 \, \alpha \alpha}^{\eta} \sigma_{\alpha}$.

We use the variables $g_{0t}^{\eta}$, $g_{0l}^{\eta}$ as adjustable parameters to reproduce the $g$-factors of the bulk PbS and PbSe, calculated in the tight-binding model (see e.g. Supplementary information of Ref.~\citenum{Kirstein22}), from the following relations
\begin{equation}
  \label{eq:gg_bulk_iso}
  g_{l(t),\text{bulk}}^{\eta} = g^{\eta}_{0l(t)} \pm \frac{4 P^2}{E_g m_0}\,,
\end{equation}
where the sign of the second term is positive for $\eta=c$ and negative for $\eta=v$.
The values $g_{l(t),\text{bulk}}^{\eta}$ and $P$ are given in Table~\ref{tb:kp_params}.
Note that in Ref.~\citenum{Dimmock64} contributions of remote bands to the $g$-factors were omitted, as the second term in Eqs.~\eqref{eq:gg_bulk_iso} ($13.1$ for PbS and $33.4$ for PbSe, by the absolute value) prevails in determining the bulk $g$-factors. However, in NCs, the $g$-factors are renormalized (in the first order, as a result of the quantum confinement energy being added to the band gap) and the contributions of remote bands become important.

The matrix elements of the Hamiltonian~\eqref{eq:HB_iso} between conduction (valence) band states $\ket{c,F_z}\equiv\ket{\frac12,+1,0;F_z}$ ($\ket{v,F_z}\equiv\ket{\frac12,-1,0;F_z}$), Eq.~\eqref{eq:QD_states}, can be calculated explicitly. They are reduced to one-dimensional integrals containing the radial functions~\cite{Note1}.
Linear-in-${\bf B}$ terms give
the values of the renormalized $g$-factor in the conduction band
\begin{multline} \label{eq:g_cb_kp}
  g_{t,l}^c =  
  \int\limits_0^R \mathrm{d} r r^2 
  \left( 
  {g_{0t,l}^c }  f_{0+}^2(r)
   -\frac{8P }{3\hbar} rf_{0+}(r)g_{1+}(r)  \right. \\
  \left.-\left[\frac{g^v_{0t,l} }3 + \frac{8\alpha_vm_0}{3\hbar^2}\right] g_{0+}^2(r) \right)
\end{multline}
and in the valence band
\begin{multline}
  \label{eq:g_vb_kp}
   g_{t,l}^v =  
\int\limits_0^R \mathrm{d}r r^2 \left( 
  {g^v_{0t,l} }  g_{0-}^2(r)
  -\frac{8P }{3\hbar} rf_{1-}(r)g_{0-}(r)   \right.   \\
  \left.-\left[\frac{g^c_{0t,l} }3 - \frac{8\alpha_cm_0}{3\hbar^2}\right] f_{1-}^2(r) \right)
 \,.
\end{multline}


\begin{table}
    \begin{tabular}{@{}lrr@{}}
    \hline\hline
      & PbS & PbSe \\
      \hline
      $E_g$ (eV) & $0.29397$ & $0.21288$ \\
      $\alpha_v\,m_0/\hbar^2$ &  $2.63535$ & $2.67380$ \\
      $\alpha_c\,m_0/\hbar^2$ &  $2.47176$ & $2.31955$ \\
      $2P^2/m_0$ (eV) &  $1.92597$ & $3.55495$ \\
      $g_{t,\text{bulk}}^v$ &  $-9.62387$ & $-24.19664$ \\
      $g_{l,\text{bulk}}^v$ &  $-9.99538$ & $-31.45314$ \\
      $g_{t,\text{bulk}}^c$ &  $10.13564$ & $ 25.99249$ \\
      $g_{l,\text{bulk}}^c$ &  $11.05293$ & $ 31.26533$ \\
      \hline\hline
    \end{tabular}
    \caption{Parameters of the isotropic \KP\ model.
    }%
\label{tb:kp_params}
\end{table}

In Figure~\ref{fig:g_lt} the results of the calculations within the \KP\ model according to equations~(\ref{eq:g_cb_kp},\ref{eq:g_vb_kp}) are shown in dashed lines. The \KP\ parameters used in calculations are given in Table~\ref{tb:kp_params}. The agreement with the tight-binding results is within the accuracy of definition of the $g$-factors extracted from the atomistic calculations. When NC shapes are varied~\cite{Note1}, then the $g$-factors obtained from the \KP\ model overlap with the distribution of $g$-factors in NCs of different shape, caused by the shape-sensitive valley mixing~\cite{Avdeev2020}. However, the shape sensitivity of the valley mixing has relatively little effect on the values of the $g$-factors. In Figure~\ref{fig:g_lt} we also present the results of the anisotropic \KP\ model described in details in Supplemental Material~\cite{Note1}. 
The only significant outcome of the anisotropic model is the non-vanishing difference in renormalizations of the longitudinal and transverse $g$-factors. However, this difference is small compared to the renormalizations themselves.

{\it Conclusions.}
To conclude, in this work we have approached calculation of $g$-factors for the carriers confined in PbS and PbSe NCs from two different standpoints. On one hand, we utilized the empirical tight-binding method to obtain $g$-factors of carrier states split at zero magnetic field by valley mixing and classified with respect to irreducible representations of the NC point group. On the other hand, we derived analytical equations for renormalizations of bulk carrier $g$-factors due to the effect of quantum confinement in a NC using a single-valley effective \KP\ Hamiltonian. We compared the outcomes of the two
calculations by mapping the tight-binding results onto a single valley. This allowed us to express $g$-factors of the carrier states, split by valley mixing and classified with respect to irreducible representations of the NC point group, in terms of the single-valley longitudinal and transverse $g$-factors and conclude that sensitivity of valley mixing to the NC shape, while significant for zero-field splittings, has relatively little effect on  carrier $g$-factors.





\begin{acknowledgments}
{\it Acknowledgments.}
The work of SVG was supported by NSF through DMR-2100248.
The work of IDA was supported by Russian Science Foundation under grant no. 22-72-00121 (numerical calculations).
The work of MON was supported by Russian Science Foundation under grant no. 19-12-00051 (analytical theory and symmetry analysis).
IDA also thanks the Foundation for Advancement of Theoretical Physics and Mathematics ``BASIS''.
\end{acknowledgments}

\bibliography{PbX}

\end{document}


\title{{\sl\textbf{Supplemental Material:}} Effective Land{\'e} factors of electrons and holes in lead chalcogenide nanocrystals}
\author{I.D.~Avdeev} 
\affiliation{Ioffe Institute, 194021 St.\,Petersburg, Russia}
\author{S.V.~Goupalov}
\email{serguei.goupalov@jsums.edu}
\affiliation{Ioffe Institute, 194021 St.\,Petersburg, Russia}
\affiliation{Department of Physics, Jackson State University, Jackson MS 39217, USA}
\author{M.O.~Nestoklon}
\affiliation{Ioffe Institute, 194021 St.\,Petersburg, Russia}

\begin{abstract}

This supporting information contains the: 
Explicit form of matrices entering Eq.(1).
Details of TB calculations. 
Comparison of results obtained for NCs of different shape.
Illustration of accuracy of effective $g$-factor extraction.
Details of calculations in isotropic and anisotropic \KP\ model.
Discussion of off-diagonal terms in effective magnetic field Hamiltonian.
Explicit form of the matrices transforming the basis functions of independent valleys to the basis of irreducible representations of the group $T_d$. 
Some mathematical expressions used in \KP\ calculations.
\end{abstract}

\maketitle

\section{Matrices entering Eq.~(1)}\label{sec:inv_matr}  

The matrices $J_{\gamma}$ ($\gamma=x,y,z$) 
are given by
\begin{subequations}
\begin{align}
\label{eq:S881}
&
J_x=
\begin{pmatrix}
  0 & \frac{\sqrt3}2 & 0 & 0 \\
  \frac{\sqrt3}2 & 0 & 1 & 0 \\
  0 & 1 & 0 & \frac{\sqrt3}2 \\
  0 & 0 & \frac{\sqrt3}2 & 0
\end{pmatrix}
\,,\\
&
J_y= 
\begin{pmatrix}
  0 & -\frac{\rmi\sqrt3}2 & 0 & 0 \\
  \frac{\rmi\sqrt3}2 & 0 &-\rmi & 0 \\
  0 & \rmi & 0 &-\frac{\rmi\sqrt3}2 \\
  0 & 0  & \frac{\rmi\sqrt3}2 & 0
\end{pmatrix}
\,,\\
& J_z=
\begin{pmatrix}
  \frac32 & 0 & 0 & 0 \\
  0 & \frac12 & 0 & 0 \\
  0 & 0 &-\frac12 & 0 \\
  0 & 0 & 0 &-\frac32
\end{pmatrix}\,.
\end{align}
\end{subequations}
They are written in the basis $(\ket{j},\ket{j-1},\ldots,\ket{-j})$  \cite{Varshalovich}.

\section{Tight-binding calculations}
We follow the procedure described in Refs.~\citenum{Avdeev17,Avdeev19,Avdeev2020} and use the extended tight-binding model \cite{Jancu98,Poddubny12} to compute the energies and wave functions of electron states in the conduction and valence bands for NCs of various shapes and classify them in accordance with irreducible representations of the point group $T_d$~\cite{Avdeev2020} (we restrict our consideration to stoichiometric NCs with no inversion center). The effect of the magnetic field is taken into account using the standard procedure of Ref.~\citenum{Graf95}. 
  
From the energy splittings induced by the external magnetic field we extract the constants in the effective Hamiltonain (1). 
NCs of cubic, octaherdal, and cuboctahedral shapes and various sizes are used in the calculations (see Refs.~\citenum{Kim21} and \citenum{Avdeev2020} for details). The effective NC diameter is calculated as a diameter of the sphere with the same volume as the NC. NCs with quasi-spherical 
shapes~\cite{Poddubny12} are obtained as a set of atoms of the bulk material located within a sphere with the diameter $D$ and the center lying halfway between a cation and an anion on a line parallel to the $\langle111\rangle$ direction.

\section{Nanocrystals of various shapes}

\begin{figure}[htp!]
  \includegraphics[width=0.9\linewidth]{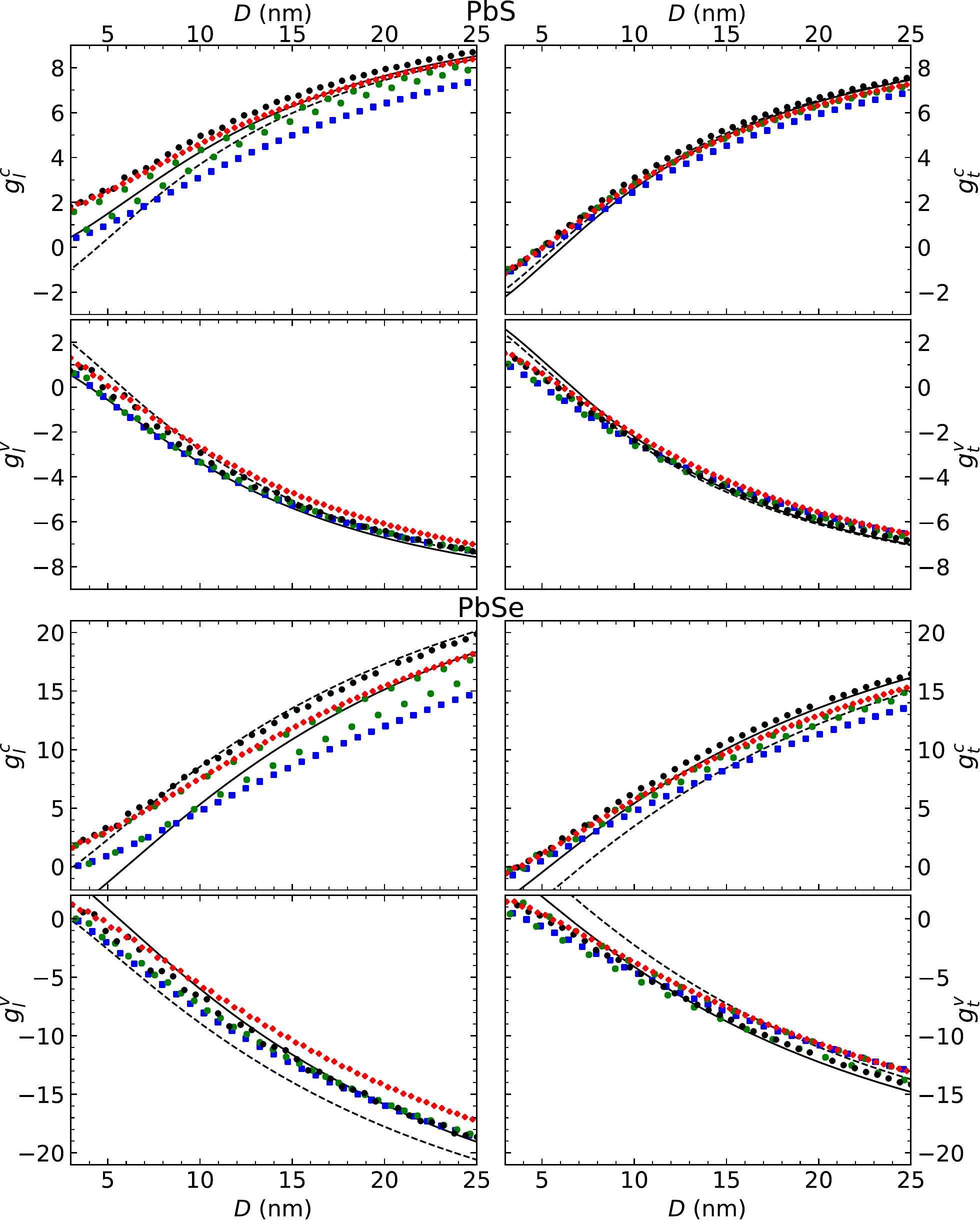}
\caption{
  Shape dependence of the effective valley $g$-factors in PbS (top four panels) and PbSe (lower four panels) NCs.
  Shape and color of the symbols represent the shapes of the NCs: cubic (blue), cuboctahedral (green), octahedral (red), and spherical (black).
  Solid (dashed) lines show outcomes of the anisotropic (isotropic) \KP\ model.
}
\label{fig:gL_shape}
\end{figure}

In the main text we show only the effective $g$-factors calculated for the quasi-spherical NCs with $T_d$ point group symmetry.
They are obtained as a set of atoms of the bulk material located within a sphere with the diameter $D$ and the center lying halfway between a cation and an anion on a line parallel to the $\langle111\rangle$ direction.\cite{Poddubny12} In figure~\ref{fig:gL_shape} we give the intra-valley $g$-factors extracted from the tight-binding calculations for NCs of various shapes: quasi-spherical, octahedral, cubic, and cuboctahedron.

\section{Accuracy of \texorpdfstring{$g$}{g}-factor determination}
\begin{figure}[htp!]
  \includegraphics[width=0.9\linewidth]{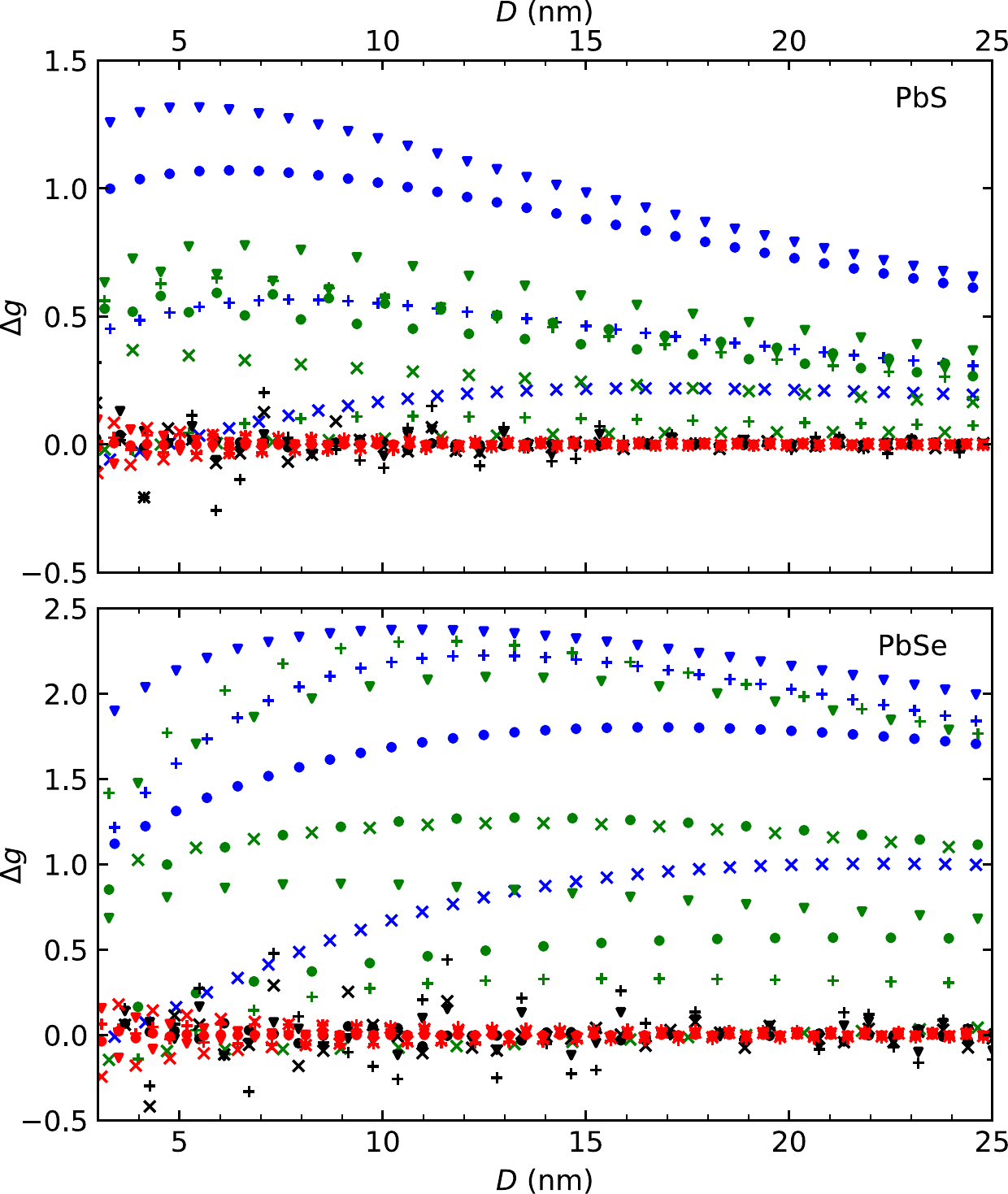}
\caption{
  Difference between the values $g_6$ ($\bullet$, +), $g_7$ ($\blacktriangledown$, $\times$) extracted from the tight-binding calculations and the values of these $g$-factors calculated after phenomenological Eqs.~(5) in conduction band ($\bullet$,$\blacktriangledown$) and valence band ($+$,$\times$).
  Color of the symbols represents the shapes of the NCs: cubic (blue), cuboctahedral (green), octahedral (red) and spherical (black).
}
\label{fig:dg_67}
\end{figure}

Since the numbers of independent constants in equations~(1) and (3) are different (four versus two), determination of the effective valley $g$-factors introduces an error.
In order to estimate it, in figure~\ref{fig:dg_67} we show the deviation of $g_6^{\eta}$ and $g_7^{\eta}$, calculated using the tight-binding method, from their counterparts obtained
with the help of $g^{\eta}_l$ and $g^{\eta}_t$.


\section{\texorpdfstring{\KP}{kp} model}

In the \KP\ calculations, we use the single-valley model of Ref.~\citenum{Kang1997}. 
Energies of the electron states with a given value of $F$, confined in a spherical NC, are found from the dispersion equations
\begin{equation}
  \label{eq:levels}
    i^{(1)}_{F-\frac{p}2}(k_- R) j_{F+\frac{p}2}(k_+ R) \rho_-  \\
 -p \, i^{(1)}_{F+\frac{p}2}(k_- R) j_{F-\frac{p}2}(k_+ R) \rho_+ =0 \,,
\end{equation}
where $p = \pm1 $ is an additional quantum number, related to the parity,
\begin{subequations}
\begin{align}
  \Lambda(E)    & = \frac{ E(\alpha_v-\alpha_c) -\frac{\hbar^2P^2}{m_0^2} - \frac{(\alpha_v+\alpha_c)E_g}2 }{2\alpha_c\alpha_v} \,, \\
  \Sigma(E)     & = \sqrt{\Lambda^2(E) + \frac{E^2-(E_g/2)^2}{\alpha_c\alpha_v}} \,,\\
  k_{\pm}(E)    & = \sqrt{\Sigma(E) \pm \Lambda(E)} \,, \\
  \rho_{\pm}(E) & = \frac{E_g/2\pm \alpha_c k_{\pm}^2(E)-E}{(\hbar P/m_0)k_{\pm}(E)} \,.
\end{align}
\end{subequations}
For each $p=\pm1$ there are positive (conduction band) and negative (valence band) roots $E_n$ of the equation \eqref{eq:levels}.
We enumerate energy levels in each band separately starting from $n=0$.

The explicit forms of the normalized radial functions in Eq.~(8) are  
\begin{align}
\label{eq:z_Fp}
\begin{split}
  f_{\ell p}(r) = & A\left[ j_{\ell} (k_+r) - \frac{j_{\ell}(k_+R)}{i^{(1)}_{\ell} (k_-R)} i^{(1)}_{\ell} (k_-r) \right]\,, \\
  g_{\ell p}(r) = &A\left[ \rho_+ j_{\ell} (k_+r) - p \rho_- \frac{j_{\ell-p} (k_+R)}{i^{(1)}_{\ell-p} (k_-R)} i^{(1)}_{\ell} (k_-r) \right] \,.
\end{split}
\end{align}
Here $j_{\ell}$ and $i_{\ell}^{(1)}$ are, respectively, the spherical Bessel and modified spherical Bessel functions of the first kind.
The normalization constant $A$ is determined from the condition
\begin{equation}
  \int\limits_0^R \left[f_{\ell p}^2(r) + g_{\ell p}^2(r)\right] r^2 \mathrm{d} r =1 \,.
\end{equation}

Since Eq.~(8) 
is written in the basis of conduction and valence band Bloch functions which are, respectively, odd and even, the partiy of the resulting wave function $\ket{F,p,n,F_z}$ is $\pi = (-1)^{F+\frac{p}2}$.

\section{Anisotropic effective mass model}
\label{sec:kp_aniso}

The effective Hamiltonian of the anisotropic model takes the form~\citenum{Kang1997} 
\begin{equation}\label{eq:H_aniso}
H_{\text{aniso}} =   \begin{pmatrix}
  \left(\frac{E_g}2 - \alpha_c^t \frac{\partial^2}{\partial x^2}- \alpha_c^t \frac{\partial^2}{\partial y^2} - \alpha_c^l \frac{\partial^2}{\partial z^2} \right) & -\frac{i\hbar}{m_0}\left( P_t \sigma_x \frac{\partial}{\partial x} +  P_t \sigma_y \frac{\partial}{\partial y} + P_l \sigma_z \frac{\partial}{\partial z}\right) \\
    -\frac{i\hbar}{m_0} \left( P_t \sigma_x \frac{\partial}{\partial x} +  P_t \sigma_y \frac{\partial}{\partial y} + P_l \sigma_z \frac{\partial}{\partial z}\right) & -\left(\frac{E_g}2 - \alpha_v^t \frac{\partial^2}{\partial x^2}- \alpha_v^t \frac{\partial^2}{\partial y^2} - \alpha_v^l \frac{\partial^2}{\partial z^2}\right)
  \end{pmatrix}
  \,,
\end{equation} 
where $P_t$ and $P_l$ are transverse and longitudinal interband momentum matrix elements and $\alpha_{c(v)}^{t(l)}$ describe
contributions to the energy dispersion from the remote bands; $x$, $y$, $z$ are the local coordinates of the valley.
The Schr\"odinger equation with this Hamiltonian can be solved numerically in the basis of the solutions of the isotropic model (8). 

An external magnetic field can be taken into account using analogs of equations~(9),~(10). The resulting Hamiltonian takes the form 
$H_{\text{aniso}} + H_{\text{aniso}}({\bf B})$, where
\begin{equation}
\label{BBB}
  H_{\text{aniso}}({\bf B}) = H'_{\text{iso}}({\bf B}) + \delta H_{P}({\bf B}) + \delta H_{\alpha}({\bf B})\,.
\end{equation}
Here  $H'_{\text{iso}}({\bf B})$ has the same form as Eq.~(10), with the electron and hole $g$-factors renormalized in accordance with\cite{Dimmock64}
\begin{align}
\begin{split}
  \label{eq:gg_bulk_aniso}
  g_{t,\text{bulk}}^c =g^c_{0t} + \frac{4 P_tP_l}{E_g m_0}
  \,,\qquad &
  g_{l,\text{bulk}}^c =g^c_{0l} + \frac{4 P_t^2}{E_g m_0}\,, 
  \\
  g_{t,\text{bulk}}^v =g^v_{0t} - \frac{4 P_tP_l}{E_g m_0}
  \,,\qquad &
  g_{l,\text{bulk}}^v =g^v_{0l} - \frac{4 P_t^2}{E_g m_0}\,.
\end{split}
\end{align}

The remaining two terms in the right-hand side of equation~(\ref{BBB}) can be written as follows:
$\delta H_{P}({\bf B}) = \delta H_{P,l}({\bf B}) + \delta H_{P,t}({\bf B})$, where
\begin{subequations}\label{eq:dHBP}
\begin{equation}
  \label{eq:dHBPl}
  \delta H_{P,l}({\bf B}) =
  \mu_B\frac{P_l-P}{\hbar}
	\sigma_z \left( B_x y -B_y x\right)
  \begin{pmatrix}
    0 & 1 \\ 1 & 0
  \end{pmatrix}
 \end{equation}
\begin{equation}
  \label{eq:dHBPt}
  \delta H_{P,t}({\bf B}) =
\mu_B \frac{P_t-P}{\hbar}
\big[
  z \left( \sigma_x B_y - \sigma_y B_x \right)
+ B_z \left( x \sigma_y - y \sigma_x \right)
\big]
  \begin{pmatrix}
    0 & 1 \\ 1 & 0
  \end{pmatrix}
  \,,
\end{equation}
\end{subequations}
and $\delta H_{\alpha}({\bf B}) = \delta H_{\alpha,l}({\bf B}) + \delta H_{\alpha,t}({\bf B})$, where
\begin{subequations}\label{eq:dHBa}
\begin{equation}
  \label{eq:dHBal}
	\delta H_{\alpha,l}({\bf B}) = -\rmi \mu_B  (B_x y - B_y x) \nabla_z \\
  \times \frac{2m_0}{\hbar^2}\begin{pmatrix}
    (\alpha_c^l-\alpha_c) & 0 \\
    0 & -(\alpha_v^l-\alpha_v)
  \end{pmatrix}
  \,,
\end{equation}
\begin{equation}
  \label{eq:dHBat}
	\delta H_{\alpha,t}({\bf B}) = -\rmi \mu_B 
	\left[ z(B_y\nabla_x-B_x\nabla_y) + \rmi B_zL_z \right]\\
  \frac{2m_0}{\hbar^2}\begin{pmatrix}
    (\alpha_c^t-\alpha_c) & 0 \\
    0 & -(\alpha_v^t-\alpha_v)
  \end{pmatrix}
  \,.
\end{equation}
\end{subequations}
Using the Wigner-Eckart theorem, each correction term, Eqs.~\eqref{eq:dHBP}, \eqref{eq:dHBa}, can be decomposed into
a one-dimensional integral containing the radial functions \eqref{eq:z_Fp} 
and an angular part.
Both these parts are computed numerically.
In practical calculations, it is enough to restrict the basis of unperturbed states by few confined states of the isotropic model with total angular momentum $F\le5/2$.

\begin{table}[tb]
  \begin{tabular}{@{}lrr@{}}
      \hline\hline
      & PbS & PbSe \\
      \hline
      $E_g $ (eV)           &  $0.29397$ &  $0.21288$ \\
      $\alpha_v^t\,m_0/\hbar^2$          &  $3.71261$ &  $3.61848$ \\
      $\alpha_v^l\,m_0/\hbar^2$          &  $0.48082$ &  $0.78444$ \\
      $\alpha_c^t\,m_0/\hbar^2$          &  $3.35927$ &  $3.00629$ \\
      $\alpha_c^l\,m_0/\hbar^2$          &  $0.69673$ &  $0.94606$ \\
      $2P_t^2/m_0$ (eV)     &  $1.67733$ &  $3.96341$ \\
      $2P_l^2/m_0$ (eV)     &  $2.42323$ &  $2.73803$ \\
      $g_{t,\text{bulk}}^v$ & $ -9.62387 $& $-24.19664$ \\
      $g_{l,\text{bulk}}^v$ & $ -9.99538 $& $-31.45314$ \\
      $g_{t,\text{bulk}}^c$ & $ 10.13564 $& $ 25.99249$ \\
      $g_{l,\text{bulk}}^c$ & $ 11.05293 $& $ 31.26533$ \\
      \hline\hline
  \end{tabular}
    \caption{The parameters of the anisotropic \KP\ model, $m_0$ is the free electron mass.}%
\label{tb:kp_params_ani}
\end{table}

\begin{figure}[htp!]
  \includegraphics[width=0.9\linewidth]{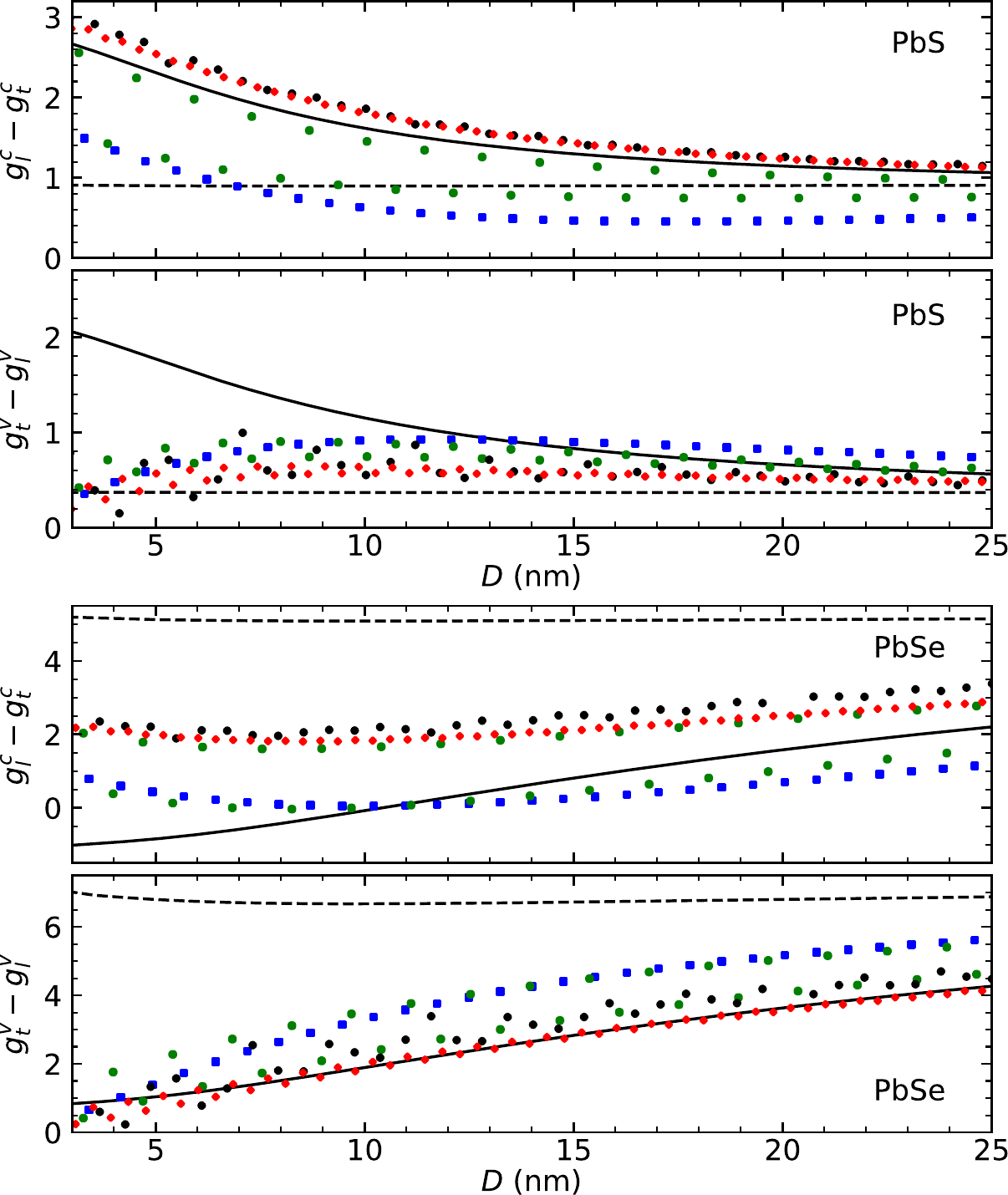}
\caption{
  Differences in the longitudinal $g^{\eta}_l$ and transverse $g^{\eta}_t$ $g$-factors in the conduction ($\eta=c$) and valence ($\eta=v$) bands of PbX quantum dots. The outcomes of the anisotropic (isotropic) \KP\ models are shown by solid (dashed) lines. By
symbols are shown the corresponding values calculated in the tight-binding method for NCs of cubic (blue), cuboctahedral (green), octahedral (red), and quasi-spherical (black) shapes.
}
\label{fig:iso_vs_aniso}
\end{figure}

While both isotropic and anisotropic \KP\ models provide a good agreement with the tight-binding calculations of the $g$-factor, only anisotropic one can account for the difference in longitudinal and transverse $g$-factors introduced by their renormalization. 
This is illustrated in Fig.~\ref{fig:iso_vs_aniso} where the differences between longitudinal and transverse $g$-factors in the conduction and valence bands, $(g^c_l-g^c_t)$ and $(g^v_t-g^v_l)$, are shown. One can see that,
in the isotropic model, the difference is almost constant.

\section{The off-diagoinal terms}
\label{app:off_diag}


The symmetry-allowed off-diagonal corrections to the Hamiltonian (1) are 
\begin{equation}\label{eq:Hinv1p}
\delta {H}_1^{\eta}({\bf B}) = 
  \mu_B{\bf B}
  \begin{pmatrix}
    0 & 0 & g_{68} {\bf  J}_{68} \\
    0 & 0 & g_{78} {\bf  J}_{78} \\
    g_{68} {\bf  J}_{68}^{\dag} & g_{78} {\bf  J}_{78}^{\dag} & 0 \\
  \end{pmatrix} \,.
\end{equation}
This form of the Hamiltonian follows from the fact that the $\Gamma_{6} \otimes \Gamma_8$ and $\Gamma_{7} \otimes \Gamma_8$ reducible representations contain the irreducible $\Gamma_4$ (pseudovector) representation of the group $T_d$.
Here ${\bf J}_{68}$ are the matrices transforming under the irreducible representation $\Gamma_4$ of the point group $T_d$ and asymmetric with respect to time inversion, written in the basis $\Gamma_{6} \otimes \Gamma_8$. Their explicit form is
\begin{subequations}
\begin{align} \label{eq:S68}
[J_{68}]_x &= 
\begin{pmatrix}
	-\frac{\sqrt3}2 & 0 & \frac12 & 0 \\
	0 & -\frac12 & 0 & \frac{\sqrt3}2
\end{pmatrix}
\,,\\
[J_{68}]_y &= -\rmi
\begin{pmatrix}
	\frac{\sqrt3}2 & 0 & \frac{1}2 & 0 \\
	0 & \frac{1}2 & 0 & \frac{\sqrt3}2
\end{pmatrix}
\,,\\
[J_{68}]_z &= 
\begin{pmatrix}
	0 & 1 & 0 & 0 \\
	0 & 0 & 1 & 0
\end{pmatrix}
\,,
\end{align}
\end{subequations}
where rows correspond to the basis functions of $\Gamma_6$ and columns correspond to the basis functions of $\Gamma_8$.
${\bf J}_{78}$ are analogous matrices in the basis $\Gamma_{7} \otimes \Gamma_8$:
\begin{subequations}
\begin{align} \label{eq:S78}
[J_{78}]_x &= 
\begin{pmatrix}
	\frac12 & 0 & \frac{\sqrt3}2 & 0 \\
	0 & -\frac{\sqrt3}2 & 0 & -\frac12
\end{pmatrix}
\,,\\
[J_{78}]_y &= \rmi
\begin{pmatrix}
	-\frac{1}2 & 0 & \frac{\sqrt3}2 & 0 \\
	0 & \frac{\sqrt3}2 & 0 & -\frac{1}2
\end{pmatrix}
\,,\\
[J_{78}]_z &= 
\begin{pmatrix}
	0 & 0 & 0 & 1 \\
	1 & 0 & 0 & 0
\end{pmatrix}
\,.
\end{align}
\end{subequations}
\begin{figure}[htp!]
  \includegraphics[width=0.9\linewidth]{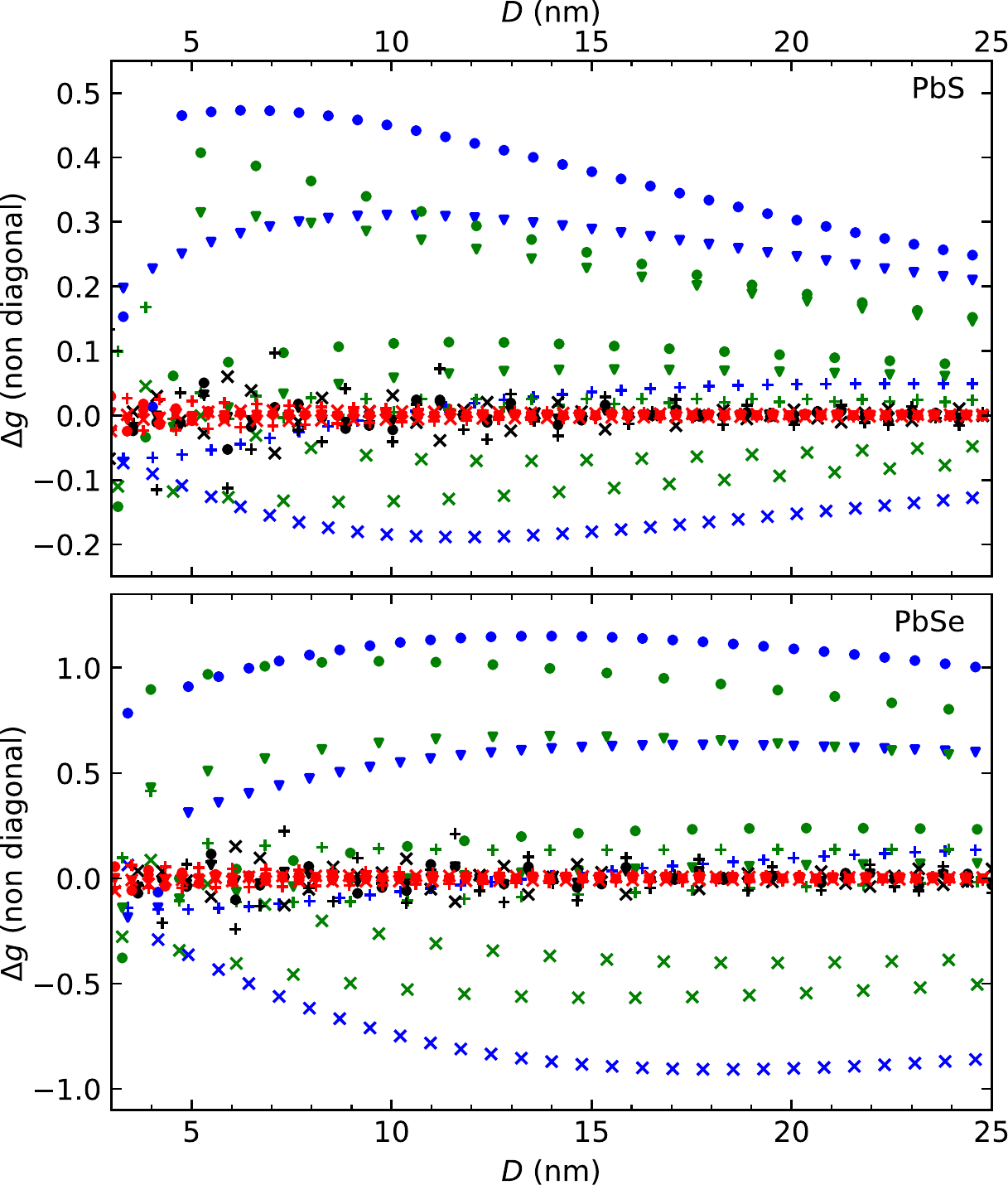}
\caption{
  Deviations of the values $g_{68}$ ($\bullet$, +) and $g_{78}$ ($\blacktriangledown$, $\times$) extracted from tight-binding calculations from phenomenological equations~\eqref{eq:g6878_glt} in conduction band ($\bullet$,$\blacktriangledown$) and valence band ($+$,$\times$).
  Color of the symbols represents the shapes of the quantum dots: cubic (blue), cuboctahedral (green), octahedral (red) and spherical (black).
}
\label{fig:dg_nd_67} 
\end{figure}

Comparison with equation~(3) yields the following relation of $g_{68}$ and $g_{78}$ to the $g$-factors in $L$ valleys:
\begin{subequations}\label{eq:g6878_glt}
\begin{alignat}{2}
  & g_{68}^c = \frac{g_l^c+g_t^c}{3\sqrt2} \,,\quad &&g_{78}^c = \frac{g_l^c-g_t^c}{3\sqrt2}\,, \\
  & g_{68}^v = \frac{g_l^c-g_t^c}{3\sqrt2} \,,\quad && g_{78}^v = -\frac{g_l^c+g_t^c}{3\sqrt2}\,.
\end{alignat}
\end{subequations}

The difference between the values $g_{68}$ and $g_{78}$ extracted from the tight-binding calculations and calculated from the phenomenological equations~\eqref{eq:g6878_glt} are shown in Figure~\ref{fig:dg_nd_67}. 
It is clear that this difference qualitatively follows the same trend as the values of the valley splittings, see Ref.~\citenum{Avdeev2020}: they are relatively large in cubic (blue), smaller in cuboctahedral (green) and even smaller in octahedral (red) and quasi-spherical (black) NCs.
In all cases, the deviations vanish with the increase of NC diameter.

\section{Transformation matrices}
\label{sec:symm_matrices}

\begin{figure}[htp!]
  \centering{\includegraphics[width=0.4\linewidth]{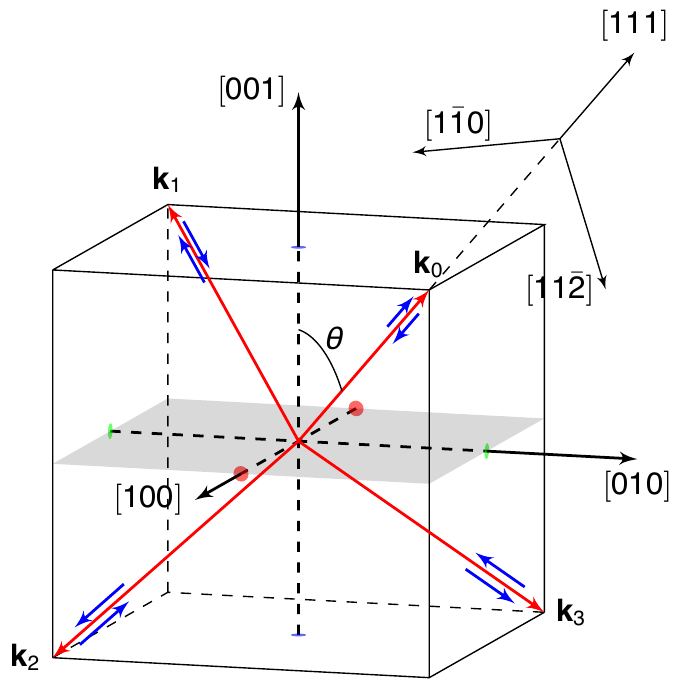}}
    \caption{The valley wave vectors ${\bf k}_{\nu}$ (red arrows) serve as directions used to define ``spin'' (total angular momentum)  
    projections $F_z = \pm 1/2$ (blue arrows) in their respective valley coordinate frames. The axes of the laboratory frame are along the $[100]$, $[010]$, and $[001]$
    directions (bold black arrows).}
    \label{fig:kks}
\end{figure}

In this Section we define transformation matrices between the basis of electronic states in independent valleys with valley-specific coordinate axes and the basis of combinations of valley states transforming according to irreducible representations $\Gamma_6$, $\Gamma_7$, and $\Gamma_8$ ($\Gamma^{\pm}_6$, $\Gamma^{\pm}_7$, and $\Gamma^{\pm}_8$) of the point group $T_d$ ($O_h$). The latter basis is defined using the laboratory reference frame with the coordinate axes $x, y,$ and $z$ along the $[100]$, $[010]$, and $[001]$ directions, respectively (see Figure~\ref{fig:kks}).

We begin with the $L_0$ valley.
In this valley the ground electron (hole) states form the basis of the $\Gamma_4^{\mp}$ irreducible representation of the $D_{3d}$ group of the wave vector ${\bf k}_{0}$. Within the isotropic \KP\ model (see Methods and Refs.~\citenum{Kang1997,Avdeev2020}), electron (hole) states are associated with pseudo-spinors (spinors) transforming under $D_{1/2}^-$ ($D_{1/2}^+$) representations of the three-dimensional orthogonal group $O(3)$.
In the other three valleys $L_{\mu}$ ($\mu=1,2,3$) we define the bases via $C_2$ rotations of the basis in the $L_0$ valley (see Figure~\ref{fig:kks} and Ref.~\citenum{Avdeev17}) as follows:
\begin{equation}
\label{eq:C2_rotations}
 L_0 \to L_1 \; \mbox{via} \; C_2^{[001]}\,,\;\;
 L_0 \to L_2 \; \mbox{via} \;  C_2^{[100]}\,,\;\;
 L_0 \to L_3 \; \mbox{via} \; C_2^{[010]}\,.
\end{equation}
The joint basis, combining all the four valley bases, represents the spin-valley basis of the star of the wave vector ${\bf k}_0$ in $T_d$ (or $O_h$) point group.

With this definition, one may find the matrices which transform the spin-valley basis to the bases of irreducible representations $\Gamma_6,\Gamma_7$, and $\Gamma_8$ of the group $T_d$. 
It is convenient to write the transformation matrices $S_{c(v)}$ as products of the three matrices:
\begin{equation}
  S_{c(v)}= D V_s U_{c(v)} \,.
 \label{Smatrix}
\end{equation}
Here the first matrix $D$ affects only spin degrees of freedom. It is a block-diagonal matrix which transforms spin states, defined in the coordinate frames of their respective valleys, to the states with certain spin projections along $z\parallel[001]$:
\begin{equation}
  D = \diag \left\{ D_0,D_1,D_2,D_3 \right\} \,.
\end{equation}
The blocks are given by
\begin{subequations}
\begin{align}\label{eq:D01}
 D_{0} &= \begin{pmatrix} 
     c_{c}e^{-\rmi\frac{\pi}8} & -\rmi s_{c}e^{ \rmi\frac{\pi}8} \\ 
   -\rmi s_{c}e^{-\rmi\frac{\pi}8} & \phantom-c_{c}e^{ \rm\rmi\frac{\pi}8} \end{pmatrix}\,,\;\;
&
 D_{1} &= \begin{pmatrix} 
     \phantom-\rmi c_{c}e^{-\rmi\frac{\pi}8} &  -s_{c}e^{ \rmi\frac{\pi}8} \\ 
      \phantom-s_{c}e^{-\rmi\frac{\pi}8} & -\rmi c_{c}e^{ \rmi\frac{\pi}8} \end{pmatrix}\,,
\\ \label{eq:D23}
 D_{2} &= \begin{pmatrix} 
      \phantom-s_{c}e^{ \rmi\frac{\pi}8} &  \phantom-\rmi c_{c}e^{-\rmi\frac{\pi}8} \\ 
     \phantom-\rmi c_{c}e^{ \rmi\frac{\pi}8} &   \phantom-s_{c}e^{-\rmi\frac{\pi}8} \end{pmatrix}\,,\;\;
&
\;\;\;\;\;
 D_{3} &= \begin{pmatrix} 
    \phantom-\rmi s_{c}e^{ \rmi\frac{\pi}8} &   c_{c}e^{-\rmi\frac{\pi}8} \\ 
    -c_{c}e^{ \rmi\frac{\pi}8} & -\rmi s_{c}e^{-\rmi\frac{\pi}8} \end{pmatrix}\,,
\end{align}
\end{subequations}
where $c_{c}=\sqrt{\frac{\sqrt3+1}{2\sqrt3}}$ and $s_{c}=\sqrt{\frac{\sqrt3-1}{2\sqrt3}}$.
These blocks were obtained assuming the following relation between spinors aligned along $z$ and along the $L_{\mu}$ valley axes
\begin{equation}
g_{0\to\mu}g_{x'}g_{z}(\up_z) = \up_{\mu}\,,\;\;\;\;\;
g_{0\to\mu}g_{x'}g_{z}(\down_z) = \down_{\mu}\,,
\end{equation}
where $g_{0\to \mu}$ is one of the three $C_2$ rotations \eqref{eq:C2_rotations} or the identity transformation (for the $L_0$ valley) and the rotation $g_{x'}g_{z}$ transforms the crystallographic axes $xyz$ to the coordinate frame of the $L_0$ valley: $x_0\parallel [1\bar10]$, $y_0 \parallel [11\bar2]$, $z_0\parallel [111]$. 
This is achieved by choosing $g_{z}$ as the clockwise rotation by angle $\pi/4$ around the $[001]$ axis and $g_{x'}$ as the clockwise rotation by the angle $\arccos(1/\sqrt3)$ around $[1\bar10]$. 
Therefore, the inverse transformation $D_{\mu}$ can be written as\footnote{Below we use the standard convention of defining the matrix of active transformation $g$ as acting on vector of basis function from the right, 
  \[ g:\;\; g({\bf e}_i) = \sum_{j} \left[D(g)\right]_{ji} {\bf e}_j   \;,\]
and the coordinates of the vectors are transformed as column vectors multiplied by the transformation matrix.}
\begin{equation}
  (\up_z,\down_z) = (\up_{\mu},\down_{\mu}) \left[D_{\frac12}(g_{0\to\mu}) D_{\frac12}(g_{x'})D_{\frac12}(g_z)\right]^{\dag}\,,
\end{equation}
where $D_{\frac12}(\omega,\bf n) = \cos{(\frac{\omega}{2})}-\rmi{\bf n} \bm{\sigma} \sin{(\frac{\omega}{2})}$ is the spin rotation matrix.

The second matrix in the right-hand side of equation~(\ref{Smatrix}), $V_s = V \otimes \mathbb 1_2$, affects only the valley degrees of freedom.
The valley wave vectors are 
${\bf k}_0 \parallel [111]$, ${\bf k}_1 \parallel [\bar1\bar11]$, ${\bf k}_2 \parallel [1\bar1\bar1]$, ${\bf k}_3 \parallel [\bar11\bar1]$, 
in agreement with~\eqref{eq:C2_rotations}.
If the spin degrees of freedom are neglected, then one can construct the spinless Bloch functions $\ket{\mu} = \rme^{\rmi {\bf k}_{\mu}r} u_{\mu}({\bf r})$ that would transform as permutations of the wave vectors ${\bf k}_{\mu}$ and form the basis of $\Gamma_1\oplus\Gamma_5$ representation of the group $T_d$. This yields the following matrix
\begin{equation}
  V = 
  \frac12
  \left(
  \begin{array}{r|rrr}
    1 & 1 & 1 & 1 \\
    1 &-1 &-1 & 1 \\
    1 & 1 &-1 &-1 \\
    1 &-1 & 1 &-1 \\
  \end{array}
  \right) \,
\end{equation}
which transforms the Bloch functions $\ket{\mu}$ as follows:
$$(\ket{0},\ket{1},\ket{2},\ket{3}) V = (\ket{S},\ket{X^+},\ket{Y^+},\ket{Z^+})\,.$$
The functions $\ket{\mu}$ and $\ket{X^+},\ket{Y^+},\ket{Z^+}$ are even since ${\bf k}_{\mu}$ and $-{\bf k}_{\mu}$ are equivalent points different by a reciprocal lattice vector.

The last matrix in the right-hand side of equation~(\ref{Smatrix}) combines the valley and spin degrees of freedom and brings the combinations of $S,X,Y,Z$ and spins $\up,\down$ to the canonical bases of the $\Gamma_6,\Gamma_7$ and $\Gamma_8$ irreducible representations of the  group $T_d$.
The matrix $U$ is different for the conduction and valence bands due to the different parities of the corresponding Bloch states.
We address this formally by adding parity to spinors.
For the conduction band, we use odd spinors $(\up^-,\down^-)$ which form the basis of $\Gamma_7$ and, therefore, the transformation takes the form
\begin{equation}
  (\ket{S},\ket{X^+},\ket{Y^+},\ket{Z^+})\otimes(\up^-,\down^-) U_c 
  = 
  (\up_6,\down_6,\up_7,\down_7,\Up_8,\up_8,\down_8,\Down_8) \,,
\end{equation}
where
\begin{equation}
  \label{eq:Uc}
  U_c = 
  \left(
  \begin{array}{cc|cc|cccc} 
     0 &                   0&  1&0  & 0                  & 0                        &0                        &0                  \\
     0 &                   0&  0&1  & 0                  & 0                        &0                        &0                  \\ \hline
     0 & \frac{\rmi}{\sqrt3}&  0&0  &-\frac{\rmi}{\sqrt2}& 0                        &\frac{\rmi}{\sqrt6}      &0                  \\
     \frac{\rmi}{\sqrt3} & 0&  0&0  & 0                  &-\frac{\rmi}{\sqrt6}      &0                        &\frac{\rmi}{\sqrt2}\\
     0 & \frac{1}{\sqrt3}   &  0&0  & \frac{1}{\sqrt2}   & 0                        &\frac{1}{\sqrt6}         &0                  \\
    -\frac{1}{\sqrt3}    & 0&  0&0  & 0                  & \frac{1}{\sqrt6}         &0                        &\frac{1}{\sqrt2}   \\
     \frac{\rmi}{\sqrt3} & 0&  0&0  & 0                  & \frac{2\rmi}{\sqrt6}&0                        &0                  \\
     0 &-\frac{\rmi}{\sqrt3}&  0&0  & 0                  & 0                        &\frac{2\rmi}{\sqrt6}&0                  \\
  \end{array}
  \right)\,.
\end{equation}

In the valence band, spinors are even $(\up^+,\down^+)$ and form the basis of $\Gamma_6$.
Therefore, in order to obtain $U_v$, the first two pairs of columns of $U_c$ should be interchanged, while, for the $\Gamma_8$ states, the following transformation should be applied to account for the difference in parities:
$$(\Up,\up,\down,\Down)\to(-\down,\Down,\Up,-\up)\,.$$
As a result,
\begin{equation}
    \label{eq:Uv}
    U_v = U_c
    \begin{pmatrix}
      0          & \mathbb1_2 & 0        & 0        \\
      \mathbb1_2 & 0          & 0        & 0        \\
      0          & 0          & 0        & \sigma_z \\
      0          & 0          &-\sigma_z & 0        \\
    \end{pmatrix}\,.
\end{equation}

We note that the derivation above is different from the procedure described in Ref.~\citenum{Avdeev2020}. First, in Ref.~\citenum{Avdeev2020} we 
arranged the basis states of the $\Gamma_6$, $\Gamma_7$, and $\Gamma_8$ representations of the point group $T_d$ in agreement with Ref.~\citenum{bookIvchenko95} while here we arranged them in the canonical form~ \cite{Varshalovich} corresponding to representations $D_{1/2}^+$, $D_{3/2}^+$ of the group $O(3)$ restricted on its subgroup $T_d \subset O(3)$. Second, in Ref.~\citenum{Avdeev2020}, the valley wave vectors were connected using the powers of $S_4$ symmetry operations while here they are related according to~(\ref{eq:C2_rotations}).

\section{Integrals used in derivation of \texorpdfstring{$g$}{g}-factors}

Eqs.~(12,13) were derived using the following integrals:
\begin{equation}
\begin{split}
\int \mathrm{d}{\bf o}\, \hat{\Omega}_{\frac12,F_z'}^{0\dag} \bm \sigma \: \hat{\Omega}_{\frac12,F_z}^{0} &= \bm \sigma_{F_z'F_z}\,,\\
\int \mathrm{d}{\bf o}\, \hat{\Omega}_{\frac12,F_z'}^{1\dag} \bm \sigma \: \hat{\Omega}_{\frac12,F_z}^{1} &= - \frac{\bm \sigma_{F_z'F_z}}3  \,,\\
\int \mathrm{d}{\bf o}\, \hat{\Omega}_{\frac12,F_z'}^{0(1)\dag} [{\bf n}_{\bf o}\times\bm \sigma]\, \hat{\Omega}_{\frac12,F_z}^{1(0)} &= \pm\frac{2\rmi}3 \bm \sigma_{F_z'F_z} \,,\\
\int \mathrm{d}{\bf o}\, \hat{\Omega}_{\frac12,F_z'}^{1\dag} {\bf L} \: \hat{\Omega}_{\frac12,F_z}^{1} &= \frac23 \bm \sigma_{F_z'F_z}\,,
\end{split}
\end{equation}
where ${\bf n}_{\bf o} = (\sin{\theta} \cos{\phi},\sin{\theta}\sin{\phi},\cos{\theta})$ and $\mathrm{d}{\bf o}=\sin{\theta} \, \mathrm{d}\theta \, \mathrm{d}\phi$.

\bibliography{PbX}